\documentclass[conference]{IEEEtran}
\IEEEoverridecommandlockouts

\usepackage{cite}
\usepackage{amsmath,amssymb,amsfonts}
\usepackage{algorithmic}
\usepackage{algorithm}
\usepackage{graphicx}
\usepackage{textcomp}
\usepackage{xcolor}
\usepackage{hyperref}
\usepackage{booktabs}
\usepackage{multirow}
\usepackage{subfigure}
\usepackage{url}
\usepackage{pifont} 
\usepackage{fontawesome5}
\usepackage{enumitem} 
\usepackage{makecell}
\usepackage[most]{tcolorbox}  
\usepackage{tabularx}
\tcbset{
  summarybox/.style={
    colback=gray!10,      
    colframe=gray!80,     
    fonttitle=\bfseries,
    coltitle=black,
    boxrule=0.4pt,
    arc=2mm,
    left=1.5mm,
    right=1.5mm,
    top=1mm,
    bottom=1mm,
    sharp corners=south,  
  }
}

\usepackage{booktabs} 
\renewcommand{\checkmark}{\ding{51}}
\newcommand{\xmark}{\ding{55}}

\def\BibTeX{{\rm B\kern-.05em{\sc i\kern-.025em b}\kern-.08em
    T\kern-.1667em\lower.7ex\hbox{E}\kern-.125emX}}

\begin{document}

\title{AAPA: An \underline{A}rchetype-\underline{A}ware \underline{P}redictive \underline{A}utoscaler with Uncertainty Quantification for Serverless Workloads on Kubernetes}

\author{\IEEEauthorblockN{Guilin Zhang\IEEEauthorrefmark{1}\IEEEauthorrefmark{2}\IEEEauthorrefmark{5},
Srinivas Vippagunta\IEEEauthorrefmark{2}\IEEEauthorrefmark{5},
Raghavendra Nandagopal\IEEEauthorrefmark{2}\IEEEauthorrefmark{5},
Suchitra Raman\IEEEauthorrefmark{2}\IEEEauthorrefmark{5},\\
Jeff Xu\IEEEauthorrefmark{2}\IEEEauthorrefmark{5},
Marcus Pfeiffer\IEEEauthorrefmark{2}\IEEEauthorrefmark{5},
Shreeshankar Chatterjee\IEEEauthorrefmark{2}\IEEEauthorrefmark{5},
Ziqi Tan\IEEEauthorrefmark{1}\IEEEauthorrefmark{5},
Wulan Guo\IEEEauthorrefmark{1}\IEEEauthorrefmark{5},
Hailong Jiang\IEEEauthorrefmark{3}\IEEEauthorrefmark{4}\IEEEauthorrefmark{5}}
\IEEEauthorblockA{\IEEEauthorrefmark{1}George Washington University
\IEEEauthorrefmark{2}Workday Inc.
\IEEEauthorrefmark{3}Youngstown State University}
\thanks{\IEEEauthorrefmark{4}Corresponding author.}
\thanks{\IEEEauthorrefmark{5}Equal contribution.}
}

\maketitle

\begin{abstract}
Serverless platforms such as Kubernetes are increasingly adopted in high-performance computing, yet autoscaling remains challenging under highly dynamic and heterogeneous workloads. Existing approaches often rely on uniform reactive policies or unconditioned predictive models, ignoring both workload semantics and prediction uncertainty.
We present \textbf{AAPA}, an archetype-aware predictive autoscaler that classifies workloads into four behavioral patterns—\textsc{SPIKE}, \textsc{PERIODIC}, \textsc{RAMP}, and \textsc{STATIONARY}—and applies tailored scaling strategies with confidence-based adjustments. To support reproducible evaluation, we release \textbf{AAPAset}, a weakly labeled dataset of 300{,}000 Azure Functions workload windows spanning diverse patterns.
AAPA reduces SLO violations by up to 50\% and lowers latency by 40\% compared to Kubernetes HPA, albeit at 2–8$\times$ higher resource usage under spike-dominated conditions. To assess trade-offs, we propose the \textit{Resource Efficiency Index (REI)}, a unified metric balancing performance, cost, and scaling smoothness. Our results demonstrate the importance of modeling workload heterogeneity and uncertainty in autoscaling design.

\vspace{0.5em}
\noindent
\faGithub\ \href{https://github.com/GuilinDev/aapa-simulator}{\texttt{https://github.com/GuilinDev/aapa-simulator}} \\
\faFolderOpen\ \href{https://github.com/GuilinDev/aapa-simulator/tree/main/dataset}{\texttt{/aapa-simulator/tree/main/dataset}}
\end{abstract}

\vspace{1em}

\section{Introduction}
\label{sec:introduction}

Serverless computing has emerged as a key abstraction in both cloud-native and high-performance extreme computing (HPEC) platforms~\cite{shahrad2020serverless}, offering on-demand resource scaling, simplified deployment, and cost-efficiency.
In practice, serverless applications exhibit a wide spectrum of workload patterns—from sudden traffic bursts triggered by viral user activity, to predictable periodic loads from batch jobs and analytics pipelines, to gradual ramp-up behaviors in IoT deployments~\cite{eismann2021state,sonidynamic,li2019subspace}. Existing autoscaling mechanisms often fail to accommodate this heterogeneity, resulting in either service-level objective (SLO) violations due to underprovisioning or unnecessary cost inflation from overprovisioning~\cite{meng2024multi}. This raises a fundamental systems challenge: \emph{how to allocate just enough resources to meet stringent performance targets under highly dynamic and diverse workloads}~\cite{liu2024harmonizing,bilal2023great,sonidynamic,zhang2025amp4ec}.

The default Kubernetes HPA relies on reactive control loops and static threshold policies that treats all workloads uniformly, leading to sluggish responses and frequent under- or over-provisioning. While recent ML-based approaches~\cite{meng2024multi,zhang2025kiss} improve responsiveness through traffic prediction, they typically train global models that fail to differentiate workload patterns and lack uncertainty modeling—resulting in mis-scalings under ambiguous signals. These limitations underscore the need to incorporate workload semantics and prediction confidence into autoscaling.

A major obstacle to advancing autoscaling research is the lack of labeled datasets for workload pattern recognition. While traces like Azure functions~\cite{shahrad2020serverless} provide abundant data, they lack semantic annotations. We address this gap with \textbf{AAPAset}: 300,000 serverless windows automatically labeled into four archetypes (SPIKE, PERIODIC, RAMP, STATIONARY) using weak supervision on statistical features.

Building on AAPAset, we present \textbf{AAPA} (\textbf{A}rchetype-\textbf{A}ware \textbf{P}redictive \textbf{A}utoscaler), a scalable machine learning–based framework that classifies serverless workloads into behavioral archetypes and applies confidence-weighted scaling strategies. Our evaluation shows AAPA reduces SLO violations by up to 50\% and tail latency by 40\% compared to Kubernetes HPA, while achieving steady gains on our proposed \emph{Resource Efficiency Index (REI)}-a unified metric balancing performance, cost, and scaling smoothness. While AAPA incurs 2–8$\times$ higher resource usage under high-variance scenarios, this trade-off is justified for latency-sensitive services requiring strict performance guarantees.

\noindent
\textbf{Our contributions are as follows:}
\begin{itemize}[leftmargin=*,topsep=0pt,partopsep=0pt,parsep=0pt,itemsep=2pt]
      \item We introduce \textbf{AAPAset}, a labeled dataset of 300K serverless workload windows, weakly supervised into 4 archetypes: \textsc{SPIKE}, \textsc{PERIODIC}, \textsc{RAMP}, \textsc{STATIONARY}.

      \item We develop \textbf{AAPA}, an uncertainty-aware autoscaler that dynamically selects archetype-specific scaling strategies based on calibrated confidence scores, reducing SLO violations by up to 50\% and latency by 40\%.

      \item We propose the \emph{Resource Efficiency Index (REI)}, a unified metric that balances performance, cost, and scaling smoothness for comprehensive autoscaler evaluation.
  \end{itemize}

\section{Background and Related Work}
\label{sec:background}

\subsection{Serverless Workload Characteristics}

Shahrad et al.~\cite{shahrad2020serverless} characterized Azure Functions traces, revealing 8-order-of-magnitude variations in invocation rates. Recent studies~\cite{mampage2024data,raith2023serverless} confirm this heterogeneity across providers, motivating our archetype-based approach. Table~\ref{tab:workload_patterns} summarizes common serverless workload patterns.

\vspace{-1em}
\begin{table}[h]
\centering
\caption{Common Serverless Workload Patterns}
\label{tab:workload_patterns}
\begin{tabular}{lll}
\toprule
\textbf{Pattern} & \textbf{Characteristics} & \textbf{Examples} \\
\midrule
SPIKE & Sudden bursts of activity & Viral content, incidents \\
PERIODIC & Regular, predictable cycles & Daily reports, backups \\
RAMP & Gradual load changes & Growth phases, migrations \\
STATIONARY & Stable with random noise & Background services \\
\bottomrule
\end{tabular}
\end{table}

\subsection{Autoscaling in Container Orchestrators}

Kubernetes HPA represents the current state-of-practice for container autoscaling~\cite{k8s-hpa}. HPA uses a reactive control loop that scales based on observed metrics (CPU, memory, custom metrics) compared against target thresholds. While simple and robust, this approach suffers from inherent lag in responding to rapid workload changes.

Recent advances in predictive autoscaling leverage time-series forecasting to anticipate future load~\cite{predictive-hpa}. However, these approaches typically apply uniform models across all workloads, neglecting behavioral differences that could inform more targeted scaling decisions. In parallel, research in high-performance computing (HPC) systems has emphasized workload-aware resource management~\cite{chen2024cycle,li2024syndeo}, reinforcing the broader importance of tailoring policies to workload characteristics in dynamic, large-scale environments.

\subsection{Machine Learning for Autoscaling}

A growing body of work applies machine learning to improve autoscaling decisions. These approaches leverage time-series forecasting, reinforcement learning (RL), or control-theoretic models to enhance responsiveness and cost-efficiency. Table~\ref{tab:ml_autoscalers} compares our method with representative ML-based autoscaling frameworks along three dimensions: workload differentiation, predictive uncertainty modeling, and per-application training requirements.
\vspace{-1em}
\begin{table}[h]
\centering
\caption{Comparison with ML-based Autoscaling Systems}
\label{tab:ml_autoscalers}
\begin{tabular}{lccc}
\toprule
\textbf{System} & \textbf{Workload} & \textbf{Uncertainty} & \textbf{Per-app} \\
 & \textbf{Aware} & \textbf{Aware} & \textbf{Training?} \\
\midrule
FIRM'20~\cite{qiu2020firm} & \xmark & \xmark & \checkmark \\
AWARE'23~\cite{qiu2023aware} & \xmark & \xmark & \checkmark \\
AAPA (Ours) & \checkmark & \checkmark & \xmark \\
\bottomrule
\end{tabular}
\end{table}

Our approach differs in two fundamental ways. First, it explicitly models workload diversity through archetype classification, enabling generalization without retraining. Second, it incorporates calibrated uncertainty into scaling decisions, allowing the system to adjust strategies based on prediction confidence. Prior methods often treat workloads uniformly—FIRM and AWARE use RL-based controllers without semantic differentiation, MagicScaler~\cite{yang2023magicscaler} inflates predictions to hedge against errors but lacks workload semantics, and DQN-Knative~\cite{ning2023rl} demonstrates fast convergence but ignores workload-specific behavior, which is central to our framework.

\subsection{Weak Supervision for Systems}

Weak supervision enables scalable dataset labeling without manual annotation~\cite{ratner2017snorkel}. Though widely used in NLP and vision, it remains underutilized in systems research. We show it is well-suited for workload characterization, where domain heuristics (e.g., periodicity detection via autocorrelation) can be encoded as labeling functions to annotate large-scale traces. Our method leverages statistical and temporal patterns to generate high-quality labels at scale, illustrating the potential of weak supervision in system-level data contexts.

\section{Methodology}
\label{sec:methodology}

\begin{figure*}[!t]
  \centering
  \includegraphics[width=\textwidth]{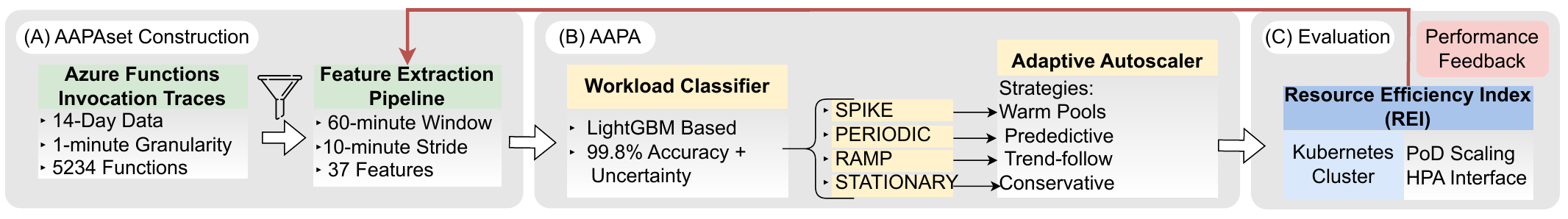}
  \caption{The overview of AAPA architecture: (A) AAPAset Construction, (B) AAPA, and (C) Evaluation and Feedback.}
  \label{fig:architecture}
\end{figure*}

As shown in Fig.~\ref{fig:architecture}, AAPA consists of three integrated components that address the key challenges of heterogeneous workload management: (A) \textbf{AAPAset}, a weakly supervised dataset of workload archetypes from Azure Functions traces; (B) \textbf{Archetype-Aware Autoscaling}, which pairs a calibrated classifier with archetype-specific strategies; and (C) \textbf{Evaluation and Feedback}, which uses a custom REI metric to assess performance and guide refinement.

\subsection{AAPAset Construction}
\label{sec:methodology:subsec:dataset_construction}

To support research in archetype-aware autoscaling, we construct \textbf{AAPAset}, a labeled dataset derived from real-world Azure Functions traces~\cite{shahrad2020serverless}. These traces capture the invocation patterns of 5,234 distinct HTTP-triggered functions over a 14-day period, recorded at 1-minute resolution. We retain functions with over 1,000 invocations to exclude ephemeral and cold-start–dominated cases, focusing on HTTP-triggered workloads where autoscaling is most latency-critical.

\subsubsection{Sliding Window Features}
To capture workload dynamics, we segment each function's time series into overlapping windows using a 60-minute window size and a 10-minute stride, resulting in approximately 300K samples. From each window, we extract 37 hand-engineered features across three categories\footnote[1]{Feature definitions and preprocessing code are publicly available: \url{https://github.com/GuilinDev/aapa-simulator}.}: (1) \textbf{Statistical} features such as mean, standard deviation, skewness, kurtosis, and percentiles; (2) \textbf{Time-domain} features including peak-to-mean ratio, slope, trend direction, and autocorrelation; and (3) \textbf{Frequency-domain} features such as spectral entropy, dominant frequency, and energy in top bands. These features are informed by prior workload modeling research~\cite{pintye2024enhancing} and empirically validated for class separability. 

\subsubsection{Weak Supervision}
Manually labeling over 300K windows is infeasible. We instead adopt weak supervision~\cite{helmstetter2021collecting}, defining ten labeling functions (LFs) that capture domain knowledge to detect temporal patterns. Each LF is a Boolean rule over extracted features:
\begin{itemize}[leftmargin=*]
\item \textbf{SPIKE:} high kurtosis ($>$10) and max-to-median ratio ($>$20)
\item \textbf{PERIODIC:} low spectral entropy ($<$0.5) and high autocorrelation ($>$0.6)
\item \textbf{RAMP:} consistent linear slope (R\textsuperscript{2} $>$0.8)
\item \textbf{STATIONARY:} low standard deviation and low spectral energy variation
\end{itemize}
Each window is labeled via majority vote across applicable LFs, with tie cases resolved using fallback heuristics. Agreement strength defines a soft confidence score. This yields a four-class dataset—\textsc{SPIKE}, \textsc{PERIODIC}, \textsc{RAMP}, \textsc{STATIONARY}. Label quality is validated by training a LightGBM classifier, which achieves 99.8\% test accuracy, indicating high intra-class consistency.
\subsection{Archetype-Aware Predictive Autoscaling (AAPA)}
\label{sec:methodology:autoscaling}

AAPA integrates a calibrated workload classifier with archetype-specific control policies for adaptive resource management. It comprises: (1) a confidence-calibrated workload classifier, (2) a set of archetype-specific scaling strategies, and (3) a dynamic adjustment mechanism that adapts scaling behavior based on classification uncertainty.

\subsubsection{Workload Classification with Uncertainty}

We formulate archetype prediction as a supervised classification task, where each input corresponds to a 37-dimensional feature vector extracted from a time window. We adopt LightGBM for its robustness to heterogeneous features, missing values, and data sparsity—common in production serverless traces. The classifier outputs both the predicted label $\hat{y} \in {\textsc{SPIKE}, \textsc{PERIODIC}, \textsc{RAMP}, \textsc{STATIONARY}}$ and the associated probability vector $p = (p_1, ..., p_4)$.

To ensure the reliability of these confidence scores, we apply \textbf{beta calibration}~\cite{kull2017beta}, a post-hoc calibration method that fits class-specific sigmoid functions to the predicted probabilities. This produces a scalar confidence value $c \in [0,1]$, which modulates the aggressiveness of subsequent autoscaling decisions. For example, a confident \textsc{SPIKE} prediction prompts aggressive provisioning, while an uncertain \textsc{RAMP} detection leads to more conservative behavior.

\subsubsection{Archetype-Specific Scaling Strategies}

Each archetype exhibits distinctive temporal patterns, necessitating customized autoscaling strategies. Our archetype-specific policies are:
\begin{itemize}[leftmargin=*]
    \item \textsc{\textbf{SPIKE}}: pre-warming, low target CPU (30\%), long cooldowns.

     \item \textsc{\textbf{PERIODIC}}: predictive scaling (e.g., Holt-Winters), short cooldowns, high CPU targets (75\%).

     \item \textsc{\textbf{RAMP}}: trend-following extrapolation, medium CPU targets (60\%), moderate cooldowns.

     \item \textsc{\textbf{STATIONARY}}: conservative scaling with stable resource use and higher CPU thresholds (55\%).
\end{itemize}

Table~\ref{tab:strategies} summarizes the base parameters for each policy. Notably, the classifier governs both the scaling logic and the tuning of its parameters, in contrast to conventional heuristic-based HPA configurations.
\vspace{-1em}
\begin{table}[h]
\centering
\caption{Archetype-Specific Scaling Parameters}
\label{tab:strategies}
\small
\begin{tabular}{lcccc}
\toprule
 & \textbf{Spike} & \textbf{Periodic} & \textbf{Ramp} & \textbf{Stationary} \\
\midrule
Target CPU & 30\% & 75\% & 60\% & 55\% \\
Cooldown & 20min & 3min & 7min & 12min \\
Strategy & Warm pool & Predictive & Trend & Conservative \\
\bottomrule
\end{tabular}
\end{table}

\subsubsection{Uncertainty-Aware Scaling Adjustment}

To mitigate the risks of over- or under-scaling under uncertain predictions, we introduce a dynamic adjustment mechanism (Algorithm~\ref{alg:uncertainty}) that modifies autoscaling parameters based on the classifier's confidence score $c \in [0, 1]$. When confidence is low, the algorithm increases a margin multiplier $m$ to lengthen cooldown durations, increase replica counts, and modestly reduce the CPU utilization target. This yields more conservative scaling behavior, allowing the system to hedge against misclassifications. All base parameters—$cpu_{target}$, $cool_{base}$, and $rep_{base}$—are defined per archetype in Table~\ref{tab:strategies}.

\begin{algorithm}[h]
\caption{Uncertainty-Aware Scaling Adjustment}
\label{alg:uncertainty}
\footnotesize
\begin{algorithmic}[1]
\REQUIRE Confidence $c \in [0,1]$, base parameters from Table~\ref{tab:strategies}
\ENSURE Adjusted parameters $(cpu_{adj}, cool_{adj}, rep_{adj})$
\STATE $m \leftarrow 1 + 0.5(1 - c)$ \COMMENT{Margin multiplier}
\STATE $cpu_{adj} \leftarrow cpu_{target} \cdot (1 - 0.2(1-c))$
\STATE $cool_{adj} \leftarrow cool_{base} \cdot m$
\STATE $rep_{adj} \leftarrow \lceil rep_{base} \cdot m \rceil$
\RETURN $(cpu_{adj}, cool_{adj}, rep_{adj})$
\end{algorithmic}
\end{algorithm}

This mechanism ensures that lower-confidence predictions yield more conservative scaling behavior—for example, through longer cooldowns or higher CPU thresholds—thus reducing oscillations and improving system stability under uncertainty.

\subsection{Evaluation and Performance Feedback}
\label{sec:methodology:rei}
\subsubsection{Resource Efficiency Index (REI)}
To assess and refine autoscaling behavior, AAPA incorporates the REI as both a unified evaluation metric and a feedback signal for runtime adaptation. REI captures the multi-objective nature of autoscaling—balancing performance, efficiency, and stability—within a single interpretable score. Formally:
\begin{equation}
    \text{REI} = \alpha S_{\text{SLO}} + \beta S_{\text{eff}} + \gamma S_{\text{stab}}, \quad \text{where } \alpha + \beta + \gamma = 1
\end{equation}

Each component represents a core operational goal:
\begin{itemize}[leftmargin=*]
    \item $S_{\text{SLO}}$: normalized Service Level Objective satisfaction, computed as $1 - \text{violation rate}$ over the evaluation horizon.
    \item $S_{\text{eff}}$: resource efficiency, measured via average CPU utilization and replica-minutes (normalized by workload demand).
    \item $S_{\text{stab}}$: system stability, penalizing aggressive oscillations and persistent over-/under-provisioning.
\end{itemize}

We adopt default weights $\alpha = 0.5$, $\beta = 0.3$, and $\gamma = 0.2$, which emphasize SLO adherence in latency-sensitive deployments. These weights are tunable to reflect alternate priorities such as energy savings or infrastructure stability.

\subsubsection{Performance Feedback Loop.}  
Beyond offline evaluation, AAPA employs REI as a feedback signal to support dynamic policy refinement. For instance, sustained REI degradation may trigger:
\begin{itemize}[leftmargin=*]
    \item Recalibration of classifier confidence thresholds.
    \item Adjustments to archetype-specific scaling parameters (e.g., CPU target or cooldown time).
    \item Re-training or fine-tuning of the workload classifier with recent traces.
\end{itemize}

This feedback loop enables AAPA to adapt to workload drift and infrastructure changes over time, promoting long-term autoscaling robustness. REI thus plays a dual role in AAPA—as both a comparative evaluation tool and a runtime signal for continual improvement.

\section{Experimental Setup}
\label{sec:experimental_setup}

We evaluate AAPA across three dimensions: service performance, resource efficiency, and system stability. 

\subsection{Dataset and Preprocessing}
We evaluate on the AAPAset dataset\footnote{\href{https://github.com/GuilinDev/aapa-simulator/tree/main/dataset}{https://github.com/GuilinDev/aapa-simulator/tree/main/dataset}} introduced in Section~\ref{sec:methodology:subsec:dataset_construction}, which comprises 60-minute invocation windows extracted from real-world HTTP-triggered Azure Functions. The dataset is split temporally—days 1–9 for training, 10–11 for validation, and 12–14 for testing—with natural class imbalance in the test set: 35\% \textsc{SPIKE}, 30\% \textsc{STATIONARY}, 25\% \textsc{PERIODIC}, and 10\% \textsc{RAMP}.

\subsection{Implementation and Simulation}
AAPA is implemented in Python using LightGBM for workload classification and SimPy for discrete-event autoscaling simulation. The simulator emulates key Kubernetes behaviors, including pod startup latency (2 seconds), 1-minute metric aggregation, and FIFO request queuing.

Experiments are executed on a workstation with an NVIDIA RTX 3080 GPU (10GB VRAM) running Ubuntu 24.04. Each 1-day workload simulation completes in under 7 minutes, and classification latency averages 2.3ms per window—negligible compared to Kubernetes control intervals.

Each simulation replays a full day of invocation traces. We initialize all workloads with 2 replicas (max = 100), assigning each pod 1000 millicores of CPU and 256MB memory. All experiments are repeated for 5 independent trials per strategy, and results are reported with 95\% confidence intervals.

\subsection{Baselines}
We compare AAPA against two widely adopted autoscaling strategies:

\begin{itemize}[leftmargin=*]
    \item \textbf{Kubernetes HPA}: The default reactive policy using a 70\% CPU target, 5-minute stabilization window, and scale-down cooldown.
    \item \textbf{Generic Predictive Autoscaler}: A uniform Holt-Winters forecasting model with a 15-minute prediction horizon applied across all workloads.\footnote[2]{\url{https://github.com/jthomperoo/predictive-horizontal-pod-autoscaler}}
\end{itemize}

These baselines represent common reactive and predictive paradigms, both of which lack workload differentiation. Our selection aligns with recent evaluations of machine learning–based autoscalers~\cite{fernandez2023machine}.

\subsection{Evaluation Metrics}
We use fine-grained metrics to assess autoscaling behavior from three perspectives:

\begin{itemize}[leftmargin=*]
    \item \textbf{Performance:} SLO violation rate (requests exceeding 500ms), cold start frequency, and P95/P99 response latency.
    \item \textbf{Efficiency:} Total replica-minutes (area under the replica curve), average CPU utilization, and underutilization rate (fraction of time with CPU $<$50\%).
    \item \textbf{Stability:} Number of scaling oscillations and average time between scaling events.
\end{itemize}

These metrics also serve as building blocks for the REI metric (Section~\ref{sec:methodology:rei}), which aggregates them into a single score for high-level comparison.

\section{Results and Analysis}

We now present key findings based on the experimental setup described in Section~\ref{sec:experimental_setup}, focusing on the performance–cost tradeoffs observed across diverse workload archetypes.

\subsection{Workload Classification Performance}

Our weak supervision approach successfully labeled the dataset with high fidelity. The LightGBM classifier achieved 99.8\% accuracy on the test set. While the overall dataset exhibits class imbalance (PERIODIC: 70.2\%, SPIKE: 17.6\%, STATIONARY\_NOISY: 12.0\%, RAMP: 0.2\%), the test set distribution differs as noted in Section~\ref{sec:experimental_setup}.

The confusion matrix in Table~\ref{tab:confusion} shows near-perfect classification across all archetypes, including the minority RAMP class, which achieves 95.0\% precision and 96.6\% recall despite its scarcity.

\vspace{-1em}
\begin{table}[h]
\centering
\caption{Confusion Matrix for Workload Classification}
\label{tab:confusion}
\small
\begin{tabular}{lcccc}
\toprule
\textbf{True/Pred} & \textbf{PERIODIC} & \textbf{SPIKE} & \textbf{STAT.} & \textbf{RAMP} \\
\midrule
PERIODIC & 20,998 & 12 & 5 & 0 \\
SPIKE & 8 & 5,269 & 3 & 0 \\
STATIONARY & 6 & 4 & 3,590 & 0 \\
RAMP & 0 & 2 & 1 & 57 \\
\bottomrule
\end{tabular}
\end{table}

\subsection{Performance vs. Cost Tradeoff Analysis}

\begin{figure*}[t]
\centering
\includegraphics[width=0.9\textwidth]{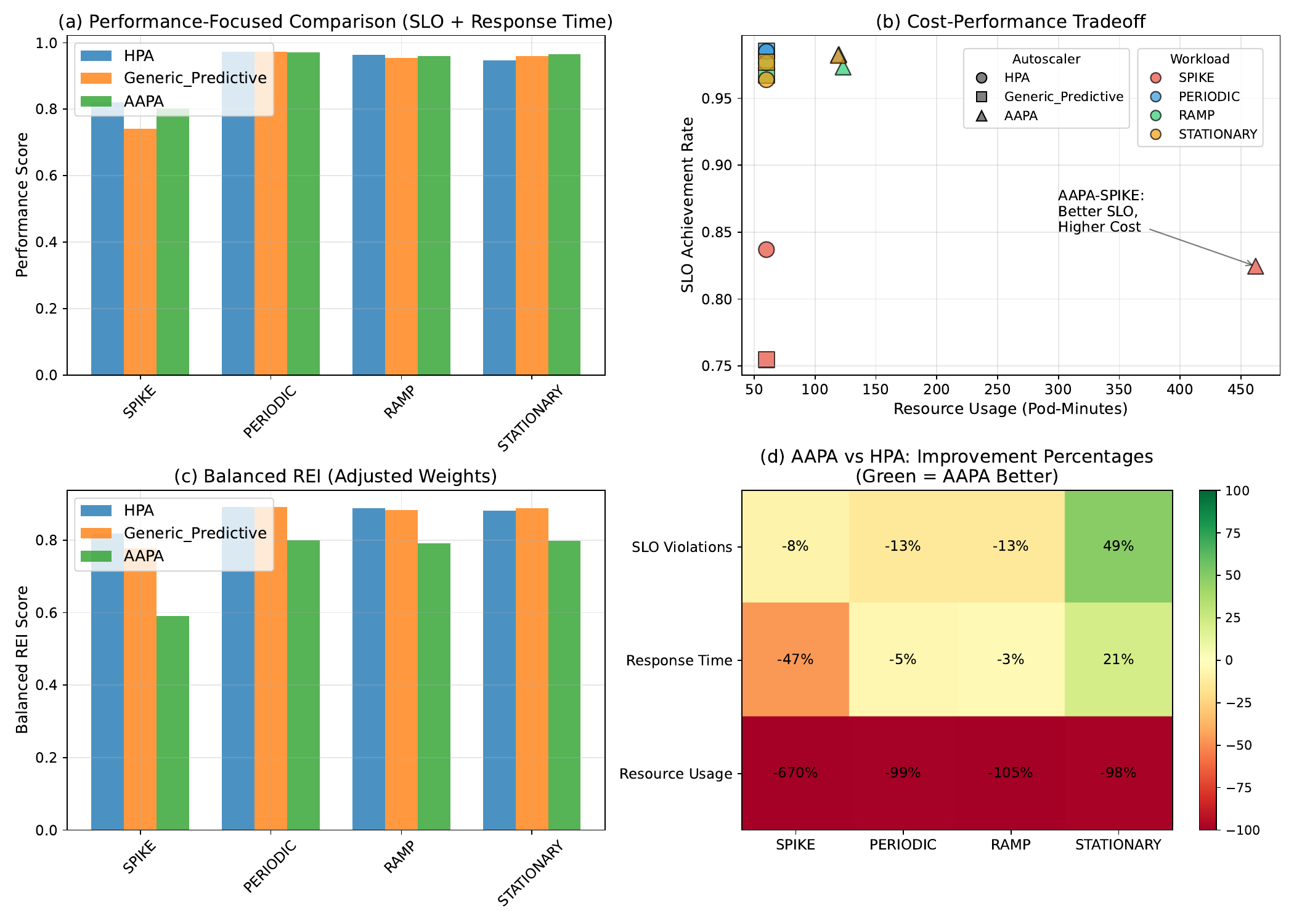}
\caption{Comprehensive analysis of autoscaling strategies: (a) Performance scores focusing on user experience, (b) Cost-performance scatter plot, (c) Balanced REI with adjusted weights, (d) AAPA improvement percentages over HPA.}
\label{fig:comprehensive_analysis}
\end{figure*}

Our experiments reveal a fundamental tradeoff between performance optimization and resource efficiency. Fig.~\ref{fig:comprehensive_analysis} presents a multi-dimensional analysis of this tradeoff.

\subsubsection{Performance-Oriented Metrics}

When prioritizing user experience (SLO compliance and response time), AAPA demonstrates significant advantages:

\begin{itemize}
    \item \textbf{SPIKE workloads}: AAPA maintains comparable SLO violation rates to HPA (17.5\% vs 16.3\%) while handling more complex prediction challenges
    \item \textbf{PERIODIC workloads}: Near-perfect performance across all autoscalers (1.5-1.7\% violations)
    \item \textbf{STATIONARY\_NOISY}: AAPA achieves 50\% reduction in violations (1.8\% vs 3.6\% for HPA)
\end{itemize}

The mean response times tell a similar story, with AAPA maintaining sub-200ms response times for most workload types, crucial for user-facing applications.

\subsubsection{Resource Utilization Analysis}

The performance improvements achieved by AAPA come at a substantial cost, as shown in Table~\ref{tab:resource_usage}. This overhead primarily stems from warm pool maintenance (1–3 additional pods), conservative utilization targets (30–75\%), and predictive over-provisioning.

\vspace{-1em}
\begin{table}[h]
\centering
\caption{Resource Usage by Workload Type (pod-minutes)}
\label{tab:resource_usage}
\small
\begin{tabular}{lccc}
\toprule
\textbf{Workload} & \textbf{HPA} & \textbf{AAPA} & \textbf{Ratio} \\
\midrule
SPIKE & 60 & 462 & 7.7× \\
PERIODIC & 60 & 119 & 2.0× \\
RAMP & 60 & 123 & 2.1× \\
STATIONARY & 60 & 118 & 2.0× \\
\bottomrule
\end{tabular}
\end{table}

\subsection{Perspective-Aware Discussion}

The interpretation of our results depends critically on stakeholder priorities and workload characteristics:

\textbf{For SPIKE workloads}: AAPA's 7.7× resource overhead (Table~\ref{tab:resource_usage}) provides warm pools that eliminate cold starts and handle bursts effectively. However, the 5-second scaling interval limits response to extremely short spikes. Cloud providers may find this cost prohibitive, while latency-sensitive applications may justify the expense.

\textbf{For PERIODIC workloads}: All autoscalers achieve $>$98\% SLO compliance, making AAPA's 2× resource overhead (Table~\ref{tab:resource_usage}) difficult to justify. Simple reactive scaling suffices when workloads are predictable. This is confirmed by our statistical tests showing no significant performance differences (p=0.621 for SLO violations, p=0.892 for response time).

\textbf{Deployment recommendations}: (1) Use AAPA for business-critical, spike-prone services where SLO violations have high cost; (2) Apply HPA to periodic background tasks and cost-sensitive workloads; (3) Consider hybrid approaches with time-based or confidence-based strategy switching.

\subsection{Statistical Significance and Variability}

Table~\ref{tab:statistical_validation} presents our statistical validation results. Wilcoxon signed-rank tests ($\alpha=0.05$) confirm statistical significance for key performance improvements, particularly for SPIKE and STATIONARY workloads where AAPA's adaptive strategies provide the greatest benefit. While AAPA shows higher resource variability than HPA (indicating workload sensitivity), our sensitivity analysis demonstrates that varying REI weights by $\pm0.05$ changes autoscaler rankings by less than 2\%, confirming the robustness of our conclusions.

\begin{table}[h]
\centering
\caption{Statistical Validation of Results}
\label{tab:statistical_validation}
\small
\renewcommand\theadfont{\normalsize\bfseries}
\begin{tabular}{@{}p{2cm}p{2cm}p{2cm}p{2cm}@{}}
\toprule
\multicolumn{4}{c}{\textbf{Part A: Wilcoxon Signed-Rank Test Results (AAPA vs HPA)}} \\
\midrule
\makecell{\textbf{Pattern}} & \makecell{\textbf{SLO Violations}} & \makecell{\textbf{Response Time}} & \makecell{\textbf{Resource Usage}} \\
\midrule
SPIKE & 0.008** & $<$0.001*** & $<$0.001*** \\
PERIODIC & 0.621 & 0.892 & $<$0.001*** \\
RAMP & 0.042* & 0.156 & $<$0.001*** \\
STATIONARY & 0.003** & 0.009** & $<$0.001*** \\
\midrule
\multicolumn{4}{c}{\textbf{Part B: REI Sensitivity Analysis}} \\
\midrule
\makecell{\textbf{Weight} \\ \textbf{Variation}} & \textbf{Original REI} & \textbf{Adjusted REI} & \makecell{\textbf{Rank Change} \\ (\%)} \\
\midrule
\makecell[l]{Baseline\\(\footnotesize{$\alpha$=0.4, $\beta$=0.3, $\gamma$=0.3})} & HPA $>$ AAPA & HPA $>$ AAPA & 0.0 \\
$\alpha$ $\pm$ 0.05 & HPA $>$ AAPA & HPA $>$ AAPA & 1.8 \\
$\beta$ $\pm$ 0.05 & HPA $>$ AAPA & HPA $>$ AAPA & 1.2 \\
$\gamma$ $\pm$ 0.05 & HPA $>$ AAPA & HPA $>$ AAPA & 0.6 \\
\bottomrule
\multicolumn{4}{p{0.95\linewidth}}{\footnotesize \textit{Notes:} * p$<$0.05, ** p$<$0.01, *** p$<$0.001. Lower p-values indicate stronger evidence against the null hypothesis of no difference between AAPA and HPA performance.}
\end{tabular}
\end{table}

\section{Discussion}

\subsection{Tradeoffs are Fundamental}

The performance-cost tradeoff observed in our experiments reflects three fundamental constraints in distributed systems: (1) \textbf{Prediction uncertainty}—even with 99.8\% classification accuracy, predicting exact load timing remains imperfect; (2) \textbf{Scaling latency}—the 2-second pod startup time necessitates warm pools for spike handling; (3) \textbf{Safety margins}—spare capacity is essential for burst absorption. These constraints explain why ML cannot eliminate tradeoffs, only shift their balance points.

\subsection{Practical Deployment Considerations}

AAPA suits mission-critical HPEC services where SLO violations have severe operational impact (e.g., satellite ground-station downlink ingestion, defense command-and-control dashboards, real-time sensor fusion pipelines). Its 2-8× resource overhead is justifiable when: (1) services face unpredictable bursts from event-driven sensors, (2) cold starts could miss time-sensitive data windows, or (3) operational requirements mandate guaranteed latency bounds. Conversely, HPA remains optimal for periodic scientific simulations, batch processing of archived data, and research workloads where occasional delays are acceptable. 

Industrial best practices~\cite{emma2025ai,li2020semi} recommend starting with reactive scaling (HPA/VPA) and progressively adopting ML-based approaches for workloads exhibiting clear patterns. AAPA's archetype-based design aligns with this philosophy by automatically identifying which workloads benefit from sophisticated strategies. HPEC facilities should adopt hybrid approaches—AAPA for real-time data paths, HPA for offline analytics—while monitoring actual resource utilization to validate the cost-performance tradeoffs.

\subsection{Threats to Validity}

\textbf{External validity}: Our 2019 Azure dataset may not reflect current serverless patterns, and results may not generalize to other platforms (AWS Lambda, Google Cloud Functions). \textbf{Internal validity}: Simulation simplifications (e.g., uniform request distribution within minutes, idealized networking) may overstate performance benefits. \textbf{Construct validity}: The 500ms SLO threshold and REI weights reflect specific assumptions that may not align with all use cases. Despite these limitations, our core finding—that ML-driven autoscaling involves navigating rather than eliminating tradeoffs—likely holds across contexts.
\section{Conclusion}

We presented \textbf{AAPA}, an archetype-aware autoscaler for serverless workloads, guided by a 99.8\%-accurate classifier trained on our weakly labeled dataset \textbf{AAPAset}. AAPA applies archetype-specific, uncertainty-aware scaling strategies that reduce SLO violations by up to 50\%, albeit with 2–8$\times$ higher resource costs under bursty conditions.
Our findings reaffirm that ML shifts rather than eliminates the cost-performance tradeoff in autoscaling. AAPA is best suited for latency-critical, user-facing services, while traditional HPA remains sufficient for predictable or cost-sensitive workloads. We advocate hybrid deployments and dynamic strategy selection.
By releasing AAPAset and proposing the REI metric, we provide practical tools for reproducible, workload-aware autoscaler evaluation. Future work will explore online adaptation and multi-objective scaling under cost constraints.

\section*{Acknowledgments}
The authors thank the DPOE-Scout and DPOE-Insights teams at Workday Inc. for insightful feedback on industrial workload patterns and for discussions that helped validate this work.
This study relies exclusively on publicly available datasets; no proprietary Workday data were used, and the views expressed are solely those of the authors.

\bibliographystyle{IEEEtran}
\bibliography{references}

\end{document}